\documentclass[aps,prl,preprint,superscriptaddress]{revtex4-1}
\usepackage[pdftex]{graphicx}

\begin{document}


\title{Constraining the 6.05 MeV 0$^+$ and 6.13 MeV 3$^-$ cascade transitions in
the $^{12}$C($\alpha$,$\gamma$)$^{16}$O reaction using the Asymptotic Normalization Coefficients}


\author{M.L. Avila}
\email[]{mavila@anl.gov}

\altaffiliation{Physics Division, Argonne National Laboratory, Argonne IL 60439, USA}
\affiliation{Department of Physics, Florida State University, Tallahassee, FL 32306, USA}

\author{G.V. Rogachev}
\email{rogachev@tamu.edu}
\affiliation{Department of Physics\&Astronomy and Cyclotron Institute, Texas A\&M University, College Station, 
TX 77843, USA}

\author{E. Koshchiy}
\affiliation{Department of Physics\&Astronomy and Cyclotron Institute, Texas A\&M University, College Station, 
TX 77843, USA}

\author{L.T. Baby}
\affiliation{Department of Physics, Florida State University, Tallahassee, FL 32306, USA}

\author{J. Belarge}
\affiliation{Department of Physics, Florida State University, Tallahassee, FL 32306, USA}

\author{K.W. Kemper}
\affiliation{Department of Physics, Florida State University, Tallahassee, FL 32306, USA}

\author{A.N. Kuchera}
\altaffiliation{National Superconducting Cyclotron Laboratory, Michigan State University, East Lansing, 
MI 48824, USA}
\affiliation{Department of Physics, Florida State University, Tallahassee, FL 32306, USA}

\author{A. M. Mukhamedzhanov}
\affiliation{Department of Physics\&Astronomy and Cyclotron Institute, Texas A\&M University, College Station, 
TX 77843, USA}

\author{D. Santiago-Gonzalez}
\altaffiliation{Department of Physics and Astronomy, Louisiana State University, Baton Rouge, LA 70803, USA}
\affiliation{Department of Physics, Florida State University, Tallahassee, FL 32306, USA}

\author{E. Uberseder}
\affiliation{Department of Physics\&Astronomy and Cyclotron Institute, Texas A\&M University, College Station, 
TX 77843, USA}

\date{\today}

\begin{abstract}
\begin{description}
\item[Background] The $^{12}$C$(\alpha,\gamma)^{16}$O reaction plays a fundamental role in astrophysics and needs to be known
with accuracy better than 10\%. Cascade $\gamma$ transitions through the excited states of $^{16}$O are contributing to 
the uncertainty. 
\item[Purpose] To constrain the contribution of the 0$^+$ (6.05 MeV) and 3$^-$ (6.13 MeV) cascade transitions. 
\item[Method] The $\alpha$ Asymptotic Normalization Coefficient for these states were measured using the $\alpha$-transfer
reaction $^{12}$C$(^6$Li,$d)^{16}$O at sub-Coulomb energies. 
\item[Results] The 0$^+$ and 3$^-$ cascade transitions contribution was found to be 1.96$\pm$0.3 keV b and 0.12$\pm$0.04 keV b for 
destructive interference of the direct and resonance capture and 4.36$\pm$0.45 keV b and 1.44$\pm$0.12 keV b for constructive 
interference respectively. 
\item[Conclusions] The combined contribution of the 0$^+$ and 3$^-$ cascade transitions to the reaction at 300 keV does 
not exceed 4\%. Significant uncertainties have been dramatically reduced.
\end{description}
\end{abstract}


\maketitle


The radiative capture of $\alpha$-particles on $^{12}$C plays a fundamental role in astrophysics. The 
$^{12}$C($\alpha,\gamma$)$^{16}$O reaction is activated during the helium burning stages of stellar evolution.
This reaction becomes important when the triple-$\alpha$ reaction, dominant during the initial
stage of helium burning, produces significant abundance of carbon. 
At temperatures that correspond to helium burning the $\alpha$-capture is most efficient at energies near 300 keV
(Gamow energy). The $^{12}$C($\alpha,\gamma$)$^{16}$O reaction cross section at this energy determines the relative 
abundance of $^{12}$C/$^{16}$O in the stellar core, which is crucial for the later stellar burning stages, 
in particular, for the rates of the reaction $^{16}$O$(\alpha,\gamma)^{20}$Ne.
This, in turn, has important implications for the sequence of later quiescent and 
explosive burning stages in stars, including nucleosynthesis and production of long-lived radioactive isotopes,
such as $^{26}$Al, $^{44}$Ti and $^{60}$Fe in core collapse supernova \cite{Clar10}. It also has direct 
influence on the composition of white dwarfs, and therefore plays an important role in the type Ia supernova
ignition process (see Ref. \cite{Wies12} and references therein). 

Significant progress in constraining the $^{12}$C($\alpha,\gamma$)$^{16}$O reaction rate has been 
achieved over the last 40 years, however, the astrophysically required precision of better than 10\% 
\cite{Weav93} is still out of reach. This is because direct measurement of radiative $\alpha$-capture 
reaction on $^{12}$C at 300 keV is unfeasible (cross section is $\sim$10$^{-17}$ b) and extrapolations
from higher energy measurements have to be used. However, extrapolations are difficult because there
are no resonances near 7.5 MeV excitation energy in $^{16}$O that can dominate the cross section 
(300 keV above the $\alpha$-decay threshold) and the $\alpha$-capture process is determined by the mixture 
of ground state and cascade transitions. It was assumed in the past that the ground state transition through
the tails of sub-threshold states and above threshold resonances plays a dominant role and that cascade 
transitions are relatively 
unimportant. This assumption was called into question in \cite{Mate06} where the S-factor at 300 keV for the 
0$^+$ state at 6.05 MeV cascade transition was determined to be 25$^{+16}_{-15}$ keV b. This is comparable to 
the $E2$ transition to the ground state (53$^{+13}_{-18}$ keV b) \cite{Buch06} and corresponds to 15\% of the 
total S-factor. Very different conclusions were made in Refs. \cite{Schu11,Schu12}, where the upper limit for 
the transition was set at $<$1 keV b. Both measurements were performed at higher energies ($>$2 MeV) using 
recoil separators and the results were extrapolated down to 300 keV. The discrepancy is caused mostly by 
different extrapolation approaches but it is also due to a 50\% lower cross section for the 6.05 MeV transition 
measured in \cite{Schu11,Schu12} as compared to \cite{Mate06}. This discrepancy causes significant uncertainty 
for the $^{12}$C($\alpha$,$\gamma$)$^{16}$O reaction rate. In Ref. \cite{Schu12} the contribution of another 
cascade transition, the 3$^-$ at 6.13 MeV, was determined to be negligibly small (0.3 keV b). The main 
goal of this letter is to constrain the 6.05 MeV 0$^+$ and 6.13 MeV 3$^-$ cascade transitions using an 
independent technique.

It has been shown that reliable constrains on direct proton capture transitions can be obtained if one 
determines the proton Asymptotic Normalization Coefficient (ANC) of the corresponding state \cite{Mukh97}.
A large number of proton-capture reactions have been investigated this way and results were benchmarked
against the direct measurements (see recent review paper and references therein \cite{Tribb14}). 
Application of the ANC technique for $\alpha$-capture reactions
was pioneered in Ref. \cite{Brun99}, where $\alpha$ ANCs for the $2^+$ and $1^-$  states at 6.92 and 7.12 MeV 
in $^{16}$O were measured using the sub-Coulomb $\alpha$-transfer reactions $^{12}$C($^6$Li,$d$)$^{16}$O and
$^{12}$C($^7$Li,$t$)$^{16}$O. The advantage of using sub-Coulomb energies for $\alpha$-transfer reactions is 
that the extracted ANCs are practically independent of the optical model potentials. Extracting the ANC instead 
of the spectroscopic factor eliminates uncertainties associated with the shape of the cluster form factor 
potential and the number of nodes of the cluster wave function. Therefore, results of these measurements are 
nearly model independent and do not require any additional normalization as long as the reaction mechanism is 
dominated by peripheral single-step $\alpha$-capture. This experimental approach has previously been used to 
investigate the $^{13}$C($\alpha$,$n$)$^{16}$O and $^{14}$C($\alpha$,$\gamma$)$^{18}$O reactions 
\cite{John06,John09}. More recently, a benchmark experiment was performed in Ref. \cite{Avil14} where the 
validity of the sub-Coulomb $\alpha$-transfer approach was demonstrated by measuring the ANC of the 1$^-$ state 
at 5.9 MeV in $^{20}$Ne and comparing it to the well known width of this state.

The ANCs for the $2^+$ at 6.92 MeV and $1^-$  at 7.12 MeV states in $^{16}$O have been previously measured using 
sub-Coulomb energies in Ref. \cite{Brun99} and above barrier energies in Refs. \cite{Oule12,Belh07}. However, the 
ANCs of the 0$^+$ at 6.05 MeV and 3$^-$ at 6.13 MeV states have not been measured. The 6.05-MeV 0$^+$ transition 
could not be studied in previous sub-Coulomb $^{12}$C($^6$Li,$d$)$^{16}$O measurements \cite{Brun99} because 
de-excitation $\gamma$-rays were detected and the 6.05 MeV 0$^+$ state decays by monopole ($E0$) transition
(mostly electron-positron pair creation).
The experimental technique employed in the present work does not suffer from this limitation and the ANCs 
for all relevant sub-threshold states were measured simultaneously: 6.05 MeV 0$^+$, 6.13 MeV 3$^-$, 
6.92 MeV 2$^+$ and 7.12 MeV 1$^-$.

The experiment was carried out at the John D. Fox Superconducting Accelerator Laboratory, at Florida State 
University. The cross section for $^6$Li($^{12}$C,$d$)$^{16}$O was measured at three energies of $^{12}$C beam 
(5, 7 and 9 MeV) using $^6$Li enriched targets with thickness of 35 $\mu$g/cm$^2$. The effective beam energies
in the middle of the target were 4.7 MeV, 6.75 MeV and 8.7 MeV. More details are given in Ref. \cite{Avila13,Avil14}.
For the identification of the reaction 
products two $\Delta E$-$E$ telescopes were mounted on remotely controlled rotating rings placed to the right
and left of the beam axis. Each of the $\Delta E$-$E$ telescopes was constructed with a position sensitive 
proportional counter and four pin diode 2$\times$2 cm$^2$ silicon detectors, contained in a box filled with a P10 gas
(10\% methane and 90\% Ar mixture). A Kapton foil of 7.5 $\mu$m thickness was used as the entrance window 
separating the P10 gas inside the detector from the chamber vacuum. This setup allows the measurement and identification of deuterons down to an energy of 1 MeV when 150 Torr of P10 pressure is used and also to observe
the backscattered $^6$Li ions when the pressure in the proportional counters is reduced to 50 Torr. The intensity 
of the incoming beam was measured using a Faraday cup placed at the end of the chamber.

\begin{figure}[ht]
\centering
\includegraphics[scale=0.42]{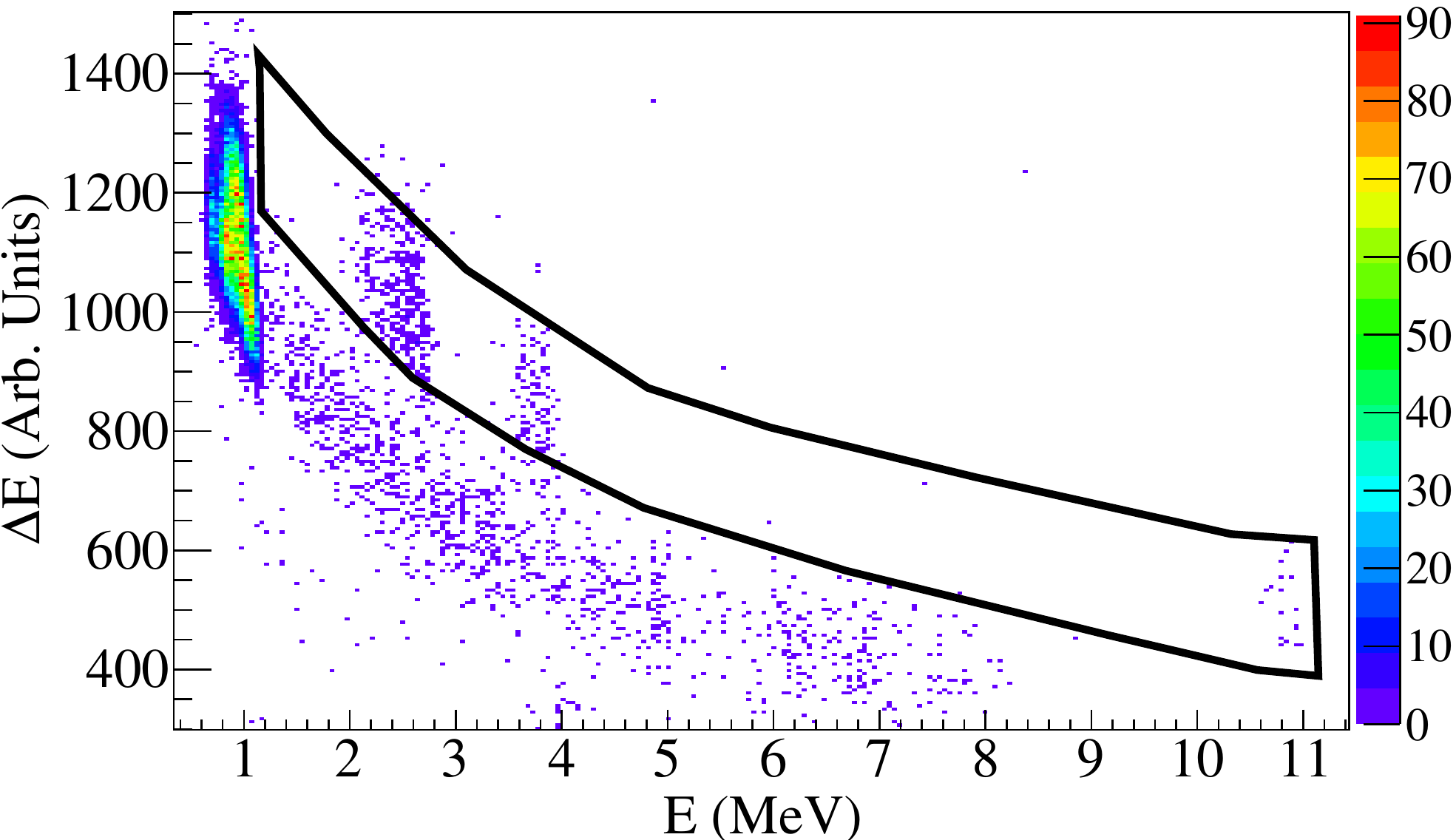}
\caption{\label{fig:EdE_all_ene} (Color online) $\Delta E$ vs $E$ scatter plot with the cut on the deuterons using the 9-MeV data 
for a pin detector at 30$^\circ$ in the laboratory frame.}
\end{figure}

The two-dimensional $\Delta E$ vs $E$ scatter plot is shown in Fig. \ref{fig:EdE_all_ene} where it can be seen 
that deuterons are clearly identified. A strong proton peak around 1 MeV is seen in Fig. \ref{fig:EdE_all_ene}. 
This peak corresponds to $^{12}$C+$p$ elastic scattering due to the hydrogen contained in the target and has a 
much higher intensity than the events of interest. The $\Delta E$ tail from these protons leaks into the deuteron 
cut and prevents deuteron identification below 1 MeV. For the 9 MeV and 7 MeV data this proton background does
not overlap with the deuterons of interest. However, for the 5 MeV data the deuterons from the 2$^+$ and 1$^-$ 
states overlap with the proton background. Therefore, only the 0$^+$ and 3$^-$ states were studied in the 
lowest energy 5-MeV dataset.

\begin{figure}[h]
\begin{center}
\includegraphics[scale=0.44]{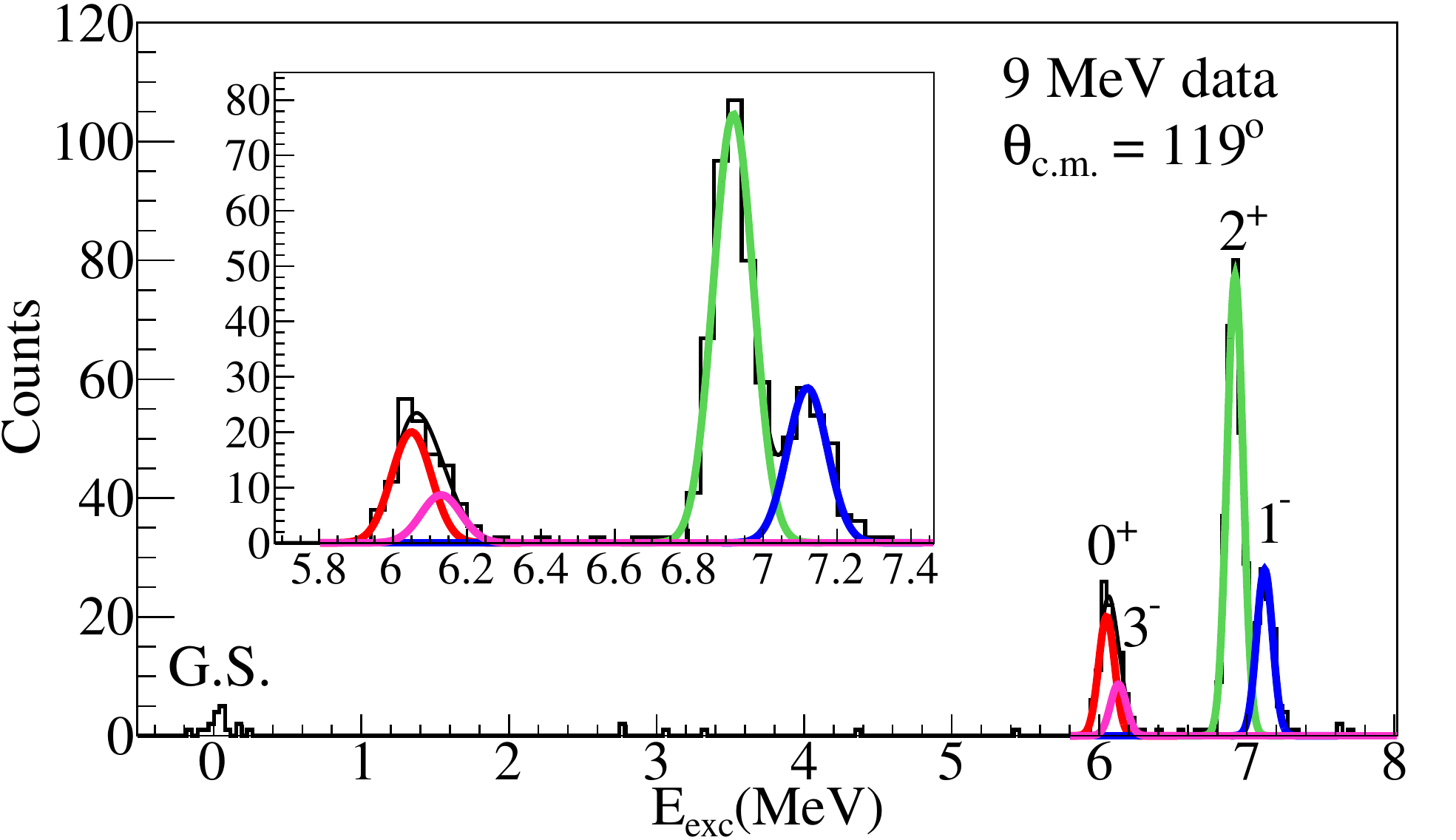}
\end{center}
\caption{\label{fig:ExcE_16O_9MeV} (Color online) Spectrum of deuterons from the $^{12}$C($^6$Li,$d$)$^{16}$O reaction. 
The $^{12}$C effective beam energy is 8.7 MeV (energy in the middle of the $^6$Li target) and the
deuteron scattering angle is 119$^{\circ}$ in the center of mass.}
\end{figure}

The spectrum of deuterons from $^{12}$C($^6$Li,$d$)$^{16}$O reaction at sub-Coulomb energy is shown in Fig. 
\ref{fig:ExcE_16O_9MeV}. As can be seen in Fig. \ref{fig:ExcE_16O_9MeV}, the energy resolution of the experiment is
good enough to resolve the 2$^+$ and 1$^-$ states that are 200 keV apart, but insufficient for clean separation 
between the 0$^+$ and 3$^-$ states that are only 80 keV apart.  The 3$^-$ state manifests itself as a shoulder 
toward the higher excitation energy visible in the peak that is dominated by the 0$^+$ state. 
We used two overlapping Gaussians fit to determine strengths of the 0$^+$ and 3$^-$ states. This 
fit has only two free parameters (amplitudes of the two Gaussians) since the excitation energies of the states
are well known and the experimental resolution is set by the widths of the resolved states (2$^+$ and 1$^-$). A
reliable fit could only be achieved for the 9- and 7-MeV dataset since the 5-MeV dataset had limited statistics. For
the 5-MeV data all the events in the 6 MeV peak were used to calculate the cross section of the 0$^+$ and 3$^-$ 
states combined. Then, the values of the ANCs and cross section for each of the states were calculated based on
the ANCs ratios of the 0$^+$ and 3$^-$ states obtained with the 9- and 7-MeV data.

\begin{figure}[h]
\begin{center}\includegraphics[scale=0.45]{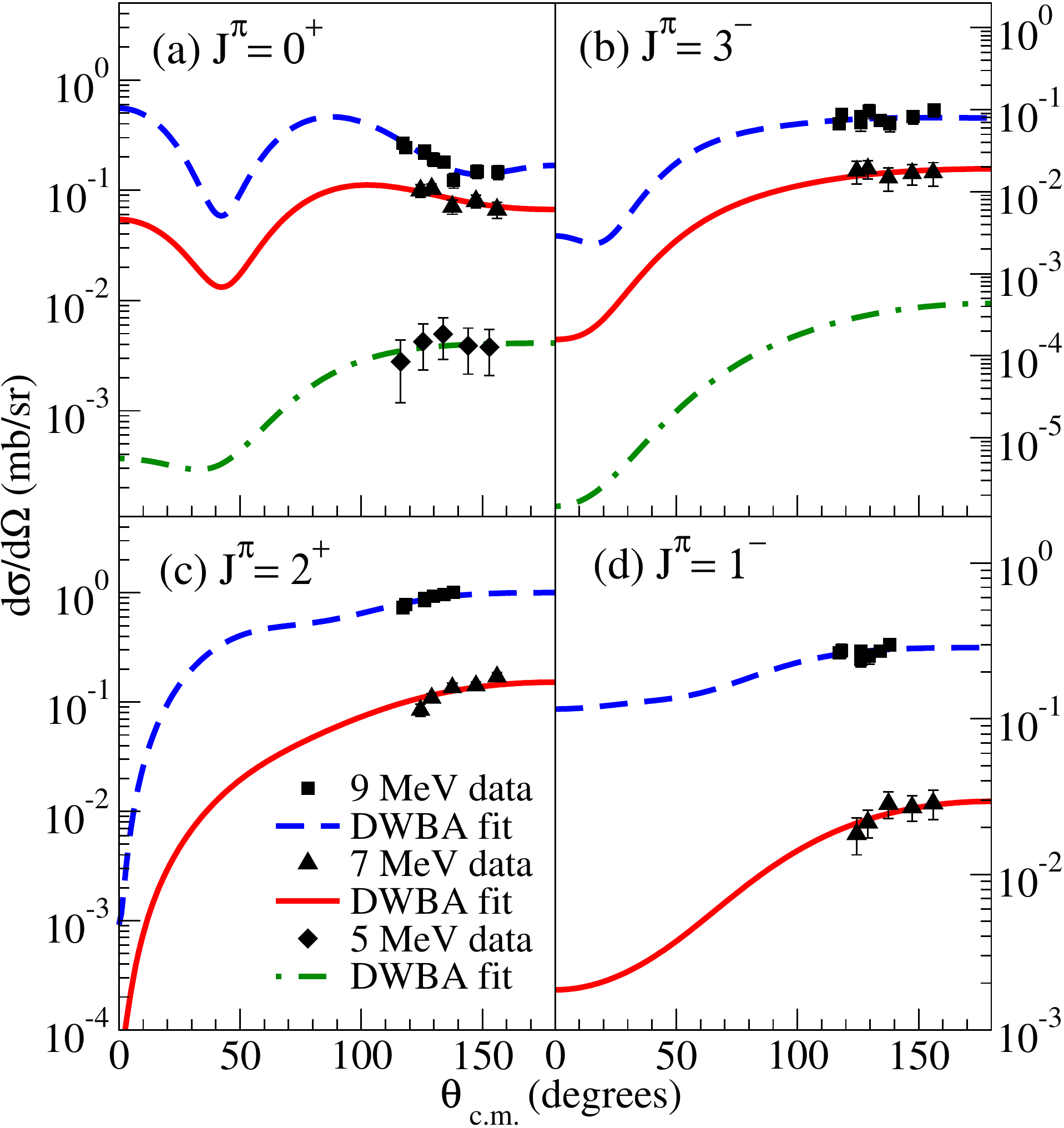}\end{center}
\caption{\label{fig:ang_dist_all} (Color online) Experimental data and DWBA cross section as a function of center-of-mass angle for
the 0$^+$ at 6.05 MeV (a), 3$^-$ at 6.13 MeV (b), 2$^+$ at 6.92 MeV (c) and 1$^-$ at 7.12 MeV (d) in $^{16}$O using 
the $^{12}$C beam energies of 5 MeV (dashed-dotted line), 7 MeV (solid line) and 9 MeV (dashed line).}
\end{figure}

Angular distributions for deuterons from the $^6$Li($^{12}$C,$d$)$^{16}$O reaction, performed at the $^{12}$C beam 
energies of 9, 7 and 5 MeV, populating 0$^+$, 3$^-$, 2$^+$ and 1$^-$ states at 6.05 MeV, 6.13 MeV, 6.92 MeV and 
7.12 MeV respectively are shown in Fig. \ref{fig:ang_dist_all} together with the corresponding DWBA calculations.
For beam energy of 5 MeV, all the data points (corrected to the 0$^+$ cross section) are shown in Fig. 
\ref{fig:ang_dist_all} (a) and only the calculated DWBA cross section is shown in Fig. \ref{fig:ang_dist_all} (b) 
for the 3$^-$ state. The computer code \textsc{fresco} (version \textsc{FRES} 2.9) \cite{Thom88} was used to perform 
finite range DWBA calculations with the full complex remnant term. The potential for $^6$Li+$^{12}$C is obtained
from Ref. \cite{Vine84}, where energy dependent parameters are obtained for energy range from 4.5 to 156 MeV 
($^6$Li beam energy). It was observed that changing the value of $V_0$ from 174 to 167 MeV produces better fit to 
the shape of the $0^+$ angular distribution. The $d$+$^{16}$O optical potential parameters were obtained from
\cite{Hint68}. The potential parameters for $\alpha$+$d$ form factor were taken from Ref. \cite{Kubo72}. By
normalizing the DWBA calculations to the experimental data and using the equations provided in Refs. 
\cite{Brun99,Mukh99} together with the known value for $^6$Li $\alpha$-ANC (($C^{^6\text{Li}}_{\alpha d}$)$^2=5.3\pm0.5$ fm$^{-1}$)
\cite{Blok93}, the ANC values for the 0$^+$ (6.05 MeV), 3$^-$ (6.13 MeV), 2$^+$ (6.92 MeV) and 1$^-$ (7.12 MeV)
states were determined. The obtained squared ANCs are shown in Table \ref{tab:16ORes} and compared to previous
measurements for the 2$^+$ (6.92 MeV) and 1$^-$ (7.12 MeV) states.

The total uncertainty of the extracted ANCs is a combination of  statistical uncertainties, normalization 
uncertainties and uncertainties in the parameters used for the DWBA calculations such as the optical potential 
parameters and the number of nodes (see Refs. \cite{Avila13,Avil14}). Due to the fact that the reaction is performed
at near and sub-Coulomb energies the uncertainty related to the optical potential parameters is small with one
exception. It was found that for the highest energy dataset (9 MeV) the angular distribution for the 6.05 MeV 0$^+$ 
state is somewhat sensitive to the optical model parameters because the exit channel ($d$+$^{16}$O) is above the
Coulomb barrier. The ANCs for each state and the corresponding uncertainties were determined for each beam energy
datasets and then combined into a single value with the corresponding statistical weights. For example, for the 0$^+$
6.05 MeV state the square of the ANCs are $\left(C^{^{16}\text{O}(0^+)}_{\alpha\textendash^{12}\text{C}}
\right)^2=(2.04\pm0.41)\times10^{6}\hspace{0.2cm}\text{fm}^{-1}$, $\left(C^{^{16}\text{O}
(0^+)}_{\alpha\textendash^{12}\text{C}}\right)^2=(2.52\pm0.50)\times10^{6}\hspace{0.2cm}\text{fm}^{-1}$
and $\left(C^{^{16}\text{O}(0^+)}_{\alpha\textendash^{12}\text{C}}\right)^2=(2.73\pm0.63)\times10^{6}
\hspace{0.2cm}\text{fm}^{-1}$ for beam energies of 9, 7 and 5 MeV respectively. This gives the average value and 
combined uncertainty of $\left(C^{^{16}\text{O}(0^+)}_{\alpha\textendash^{12}\text{C}}\right)^2=(2.43
\pm0.30)\times10^{6}\hspace{0.2cm}\text{fm}^{-1}$. The uncertainty for 9-MeV dataset is dominated by 
uncertainty of the optical potential parameters and uncertainty for the 5-MeV dataset is dominated by the statistical 
uncertainty. 

\begin{table}
\centering
\caption{\label{tab:16ORes} Squared ANCs for the 0$^+$ (6.05 MeV), 3$^-$ (6.13 MeV), 2$^+$ 
(6.92 MeV) and 1$^-$ (7.12 MeV) sub-threshold states in $^{16}$O, compared to previous measurements.}
\tabcolsep=0.005cm
\begin{tabular}{ccccc}
\hline
\hline
$\left(C^{^{16}\text{O}(0^+)}_{\alpha\textendash^{12}\text{C}}\right)^2$ &
$\left(C^{^{16}\text{O}(3^-)}_{\alpha\textendash^{12}\text{C}}\right)^2$ &
$\left(C^{^{16}\text{O}(2^+)}_{\alpha\textendash^{12}\text{C}}\right)^2$ & 
$\left(C^{^{16}\text{O}(1^-)}_{\alpha\textendash^{12}\text{C}}\right)^2$ & Ref. \tabularnewline
  (10$^6$ fm$^{-1}$)& (10$^4$ fm$^{-1}$) & (10$^{10}$ fm$^{-1}$) & (10$^{28}$ fm$^{-1}$) & \tabularnewline  	
\hline
- & - & 2.07$\pm$0.80 & 4.00$\pm$1.38 & \cite{Oule12} \tabularnewline
- & - & 1.29$\pm$0.23 & 4.33$\pm$0.84 & \cite{Brun99} \tabularnewline
- & - &1.96$^{+1.41}_{-1.27}$ & 3.48$\pm$2.0 &\cite{Belh07} \tabularnewline
2.43$\pm$0.30 & 1.93 $\pm$0.25 & 1.48 $\pm$0.16 & 4.39$\pm$0.59& This work \tabularnewline
 \hline
 \hline
\end{tabular} 
\end{table}

\begin{figure}[h]
\includegraphics[scale=0.41]{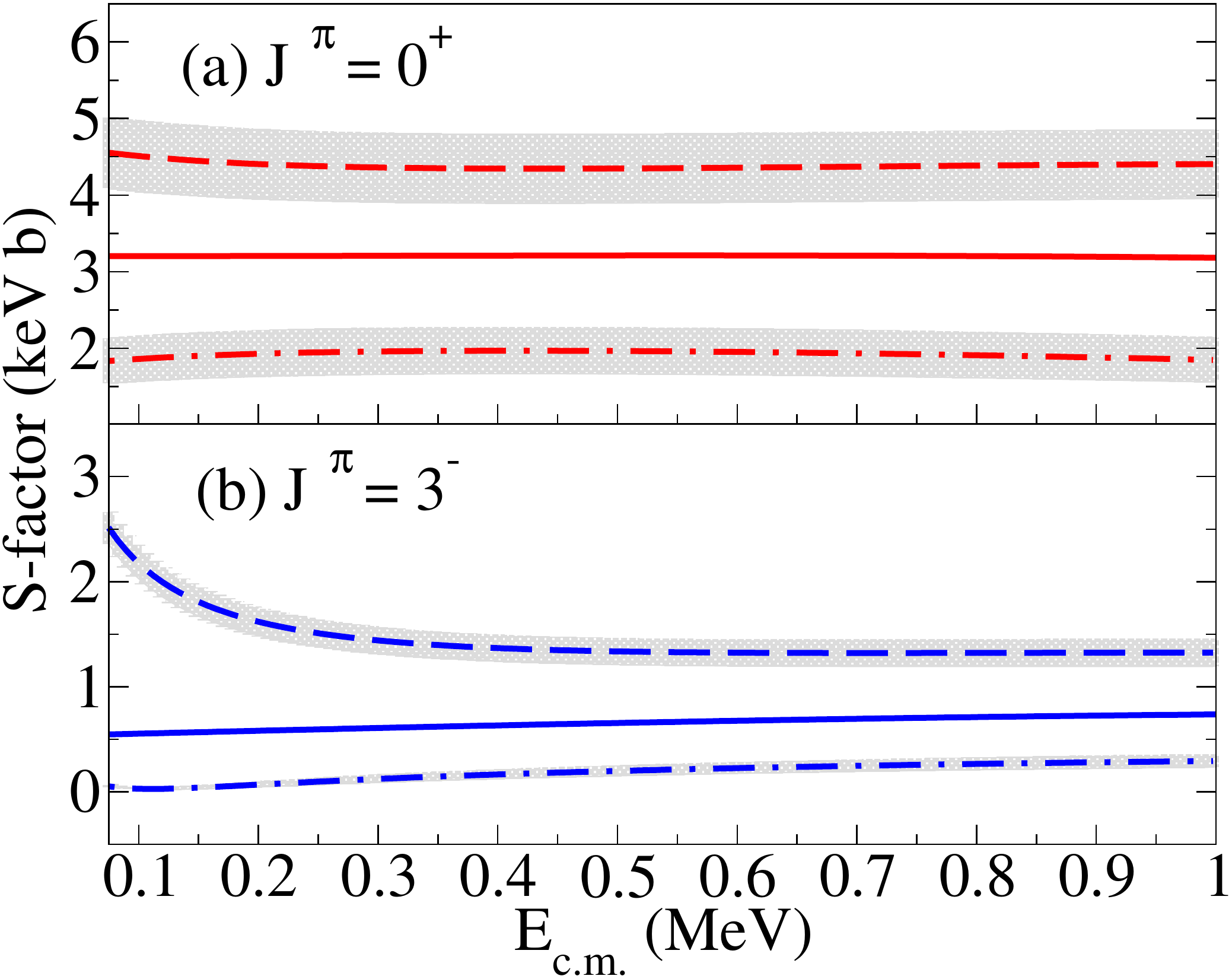}
\caption{\label{fig:S-factor_16O} (Color online) S-factor of the $^{12}$C($\alpha,\gamma$)$^{16}$O reaction
for the direct $E2$ transition to the sub-threshold 0$^+$ at 6.05 MeV (a) and for the 3$^-$ at 6.13 MeV (b)
without interference (solid line) and assuming constructive (dashed line) or destructive (dash-dotted line) interference 
with the 2$^+$ at 6.92 MeV and the 1$^-$ state at 7.12 MeV in $^{16}$O, respectively. The shaded bands correspond to one standard 
deviation uncertainties. }
\end{figure}

The ANCs for the 2$^+$ (6.92 MeV) and 1$^-$ (7.12 MeV) sub-threshold states have been 
measured in \cite{Brun99,Belh07,Oule12}. There is excellent agreement between all measurements
with sub-Coulomb $\alpha$-transfer (\cite{Brun99} and this work) providing the most precise values.
The contribution of the 2$^+$ (6.92 MeV) and 1$^-$ (7.12 MeV) sub-threshold states to the 
astrophysical S-factor have been evaluated in \cite{Brun99} using ANCs that are nearly 
identical to the results of the present work and therefore there is no need to repeat the R-matrix 
analysis already performed in \cite{Brun99}. The ANC of the 0$^+$ and 3$^-$ states have been 
measured for the first time. 
The S-factor for direct $\alpha$-capture to the 0$^+$ and 3$^-$ states was calculated using
the R-matrix formalism described in \cite{Bark91} and implemented in the code AZURE \cite{Azum10}.
The $E2$ transition dominates the direct $\alpha$-capture to the 0$^+$ and 3$^-$ states 
($E1$ and $M1$ transitions were evaluated and were found to be negligible for both cascade transitions).
The S-factor for the direct $E2$ transitions to the 0$^+$ and 3$^-$ states are shown in Fig. \ref{fig:S-factor_16O} 
(solid curve). They are completely determined by the measured ANCs with uncertainties related to the choice of channel radius 
being very small compared to the experimental uncertainties of the ANCs. 
The corresponding S-factors at 300 keV are 3.2$\pm$0.4 keV b and 0.6$\pm$0.1 keV b for direct capture to the 0$^+$ and 3$^-$
states respectively. 

The $E2$ radiative capture to the first excited state of $^{16}$O (0$^+$ at 6.05 MeV) state is contributed by the interfering
direct capture and the capture through the sub-threshold resonance 2$^+$ at 6.92 MeV. Similarly, $E2$ radiative capture 
to the second excited state of $^{16}$O (3$^-$ at 6.13 MeV) is contributed by the interfering direct capture and the capture 
through the sub-threshold resonance 1$^-$  at 7.12 MeV.
The amplitude of the capture through the sub-threshold resonance contains the product of the $\alpha$ partial width 
amplitude in the entrance channel and the radiative width amplitude for the decay of the sub-threshold resonance to
the first excited state $0^{+}$. The relative sign between these two amplitudes determines if interference is destructive or 
constructive and it cannot be determined from the experimental data presented in this letter. Therefore, we consider both
possibilities here. It is important to note that since ANCs for all sub-threshold states in $^{16}$O are fixed now, 
there is a good chance that interference sign can be determined by complete R-matrix analysis of the direct $\alpha$-capture
measurements performed at higher energies. However, this analysis is beyond the scope of this letter.

While direct capture amplitude dominates the 0$^+$ cascade transition at 300 keV, interference between the direct capture 
and the capture through the sub-threshold resonance 2$^+$ at 6.92 MeV is non-negligible and modifies the 
\textquotedblleft pure\textquotedblright direct
capture S-factor by 37\% at this energy. Therefore, the corresponding S-factor at 300 keV is either 1.96$\pm$0.3 keV b 
or 4.36$\pm$0.45 keV b for the destructive or constructive interference case respectively. The situation is more dramatic
for the 3$^-$ cascade transition. The direct capture and resonance capture through the sub-threshold 1$^-$ state at 7.12 MeV
have about equal amplitudes at 300 keV and destructive interference makes the corresponding transition very small, 
0.12$\pm$0.04 keV b. The constructive interference enhances the S-factor by a factor of 3, making it 1.44$\pm$0.12 keV b.
Fig. \ref{fig:S-factor_16O} (a) and (b) shows the S-factors for the 0$^+$ and 3$^-$ cascade transitions without 
interference (direct capture only) and constructive and destructive interference with the 2$^+$ at 6.92 MeV 
and the 1$^-$ at 7.12 MeV respectively.

This result is in obvious disagreement with \cite{Mate06} where it was found that the 0$^+$  
cascade transition contributes 25 keV b to the total S-factor. That experiment assumed $E1$ as
a dominant component for the transition. Angular distribution of $\gamma$-rays was measured in Ref. \cite{Schu11}
and it was shown that in fact $E2$ dominates.
We find that contribution from the 0$^+$ cascade transition is at least a factor of 5 smaller. 
However, we also disagree with the results of \cite{Schu12}, where the 
0$^+$ cascade transition was found to contribute only 0.3 keV b. The origin of this disagreement
is easy to point out. It turned out that the cross section for the 0$^+$ cascade transition at 300 keV 
is dominated by a single parameter - the ANC for the 0$^+$ state at 6.05 MeV. This is largely due to the  
$\alpha$-cluster nature of this state. Based on the measured ANC (and assuming a channel radius of 5.2 fm), this 
state has an $\alpha$+$^{12}$C(g.s.) spectroscopic factor of around 40\%. This is not surprising, since this state
has long been identified as a bandhead of an $\alpha$-cluster inversion doublet quasi-rotational band \cite{Cart64}.
The $\alpha$ reduced width amplitude calculated from the measured ANC (using equations provided in Ref. \cite{Mukh99}) is 
$\gamma_{6.05} = 0.48 \pm 0.08$ MeV$^{1/2}$. The reduced width amplitude used in Refs. \cite{Schu11,Schu12} for this 
state was $\gamma_{6.05} = 0.01 \pm^{0.05}_{0.01}$ MeV$^{1/2}$. This accounts for one order of magnitude difference 
in the S-factor. Such small value was based on the $^{12}$C+$\alpha$ elastic scattering data of \cite{Tisc02}. However,
an extensive R-matrix analysis of all available $^{16}$O compound nucleus reactions has
been carried out recently in Ref. \cite{deBo13}. The elastic scattering data from \cite{Tisc02} was included into the fit and 
the small $\alpha$ reduced width amplitude for the 6.05 MeV state was not confirmed. On the contrary, the ANCs 
for the 0$^+$ and 3$^- $ suggested in \cite{deBo13} (3.2$\times$10$^6$ fm$^{-1}$ and 2.3$\times$10$^4$ fm$^{-1}$, 
no uncertainties are given) are in surprisingly good agreement with the direct measurements reported here.

In summary, we have investigated the important astrophysical reaction $^{12}$C($\alpha,\gamma$)$^{16}$O through
the $\alpha$-transfer reaction $^{12}$C($^{6}$Li,$d$)$^{16}$O at sub-Coulomb energies. The ANCs for the sub-threshold
states 0$^+$ (6.05 MeV), 3$^-$ (6.13 MeV), 2$^+$ (6.92 MeV) and 1$^-$ (7.12 MeV) in $^{16}$O have been determined. 
The extracted ANCs for the 2$^+$ and 1$^-$ states are in very good agreement with previous measurements 
\cite{Brun99,Oule12,Belh07}. The ANCs of the 0$^+$ (6.05 MeV) and 3$^-$ (6.13 MeV) states were directly measured
for the first time. The uncertainties related to the contribution of the 0$^+$ (6.05 MeV) and 3$^-$ (6.13 MeV) cascade
transitions to the total S-factor at energies near 300 keV are now dramatically reduced. The cascade transitions to the 0$^+$
and 3$^-$ states were found to be determined by the interference of the E2 direct capture amplitude with the amplitude
for the sub-threshold resonance capture through the 2$^+$ state at 6.92 MeV and 1$^-$ state at 7.12 MeV respectively.
The ANCs of the corresponding states, all of which were measured in this work, determine the direct capture and 
sub-threshold resonance capture amplitudes. While interference sign is still the source of uncertainty that can be 
eliminated by performing complete R-matrix fit of higher energy experimental data for the corresponding transitions, 
the maximum contribution of the 0$^+$ and 3$^-$ cascade transitions can be determined by assuming positive interference 
in both cases. The combined contribution of these cascade transitions does not exceed 4\% of the total 
$^{12}$C($\alpha,\gamma)^{16}$O S-factor.

\begin{acknowledgments}
The authors would like to acknowledge the financial support provided by the National Science Foundation under grant 
No. PHY-456463. The authors G.V Rogachev, E. Koshchiy and A. M. Mukhamedzhanov acknowledge that this material is 
based upon their work supported by the U.S. Department of Energy, Office of Science, Office of Nuclear 
Science, under Award Number DE-FG02-93ER40773. The author G.V. Rogachev is also supported by the Welch 
Foundation (Grant No.: A-1853). A. M. Mukhamedzhanov is also supported by the U.S. Department of Energy, 
National Nuclear Security Administration, under Award Number DE-FG52-09NA29467 and by the US National Science
Foundation under Award PHY-1415656.
\end{acknowledgments}

\bibliography{myrefs}

\end{document}